\documentclass[12pt]{article}  

\usepackage{amsmath,amssymb,amsfonts}
\usepackage{longtable}

\def\mylist{\begin{list}{}{\setlength{\leftmargin}{0.5in}
               \setlength{\listparindent}{-0.5in}
               \setlength{\itemindent}{\listparindent}}}

\newcommand{\ket}[1]{|#1\,\rangle}

\newcommand{\nD}[1]{\not D}
\newcommand{\Tr}{{\rm Tr}}

\newcommand{\eps}{\epsilon}

\begin{document}

\begin{titlepage}

\title{Aharony duality and monopole operators in three dimensions}
\author{Denis Bashkirov\\ {\it California Institute of Technology}}

\maketitle

\abstract{We test dualities between three dimensional N = 2 gauge theories proposed by
Aharony in \cite{Aharony} by comparing superconformal indices of dual theories. We
also extend the discussion of chiral rings matching to include monopole operators.}

\end{titlepage}

\section{Introduction}

  An important class of dualities of four dimensional gauge theories are Seiberg dualities which relate minimally supersymmetric ${\cal N}=1$ SQCD theories with gauge group $SU(N_c)$ and $N_f$ flavors of quarks and antiquarks to $SU(N_f-N_c)$ gauge theories with $N_f$ flavors of quarks and antiquarks as well as a singlet field coupled through a superpotential. This duality has a generalization to symplectic to special orthogonal groups.

  More than a decade ago Aharony proposed a three dimensional analog of Seiberg duality. It is a duality between the infrared limits of ${\cal N}=2$ gauge theories with fundamental matter and unitary or symplectic gauge groups. Namely, an ${\cal N}=2$ supersymmetric theory with gauge group $U(N_c)$  with $N_f$ chiral fundamental multiplets and $N_f$ chiral anti-fundamental muliplets is conjectured to be dual to an ${\cal N}=2$ theory with gauge group $U(N_f-N_c)$, $N_f$ chiral fundamentals, $N_f$ chiral anti-fundamentals together with additional gauge singlet chiral fields and a superpotential. For the symplectic gauge groups the duality relates $USp(2N_c)$ gauge theory with $2N_f$ fundamental chiral fields to $USp(2N_f-2N_c-2)$ gauge theory with $2N_f$ fundamental chiral fields together with a number of gauge singlets and a superpotential.

 Another class of three-dimensional dualities for ${\cal N}=2$ and ${\cal N}=3$ theories with Chern-Simons terms was introduced by Giveon and Kutasov \cite{GK}. It was noticed by these authors that these dualities could be obtained from the Aharony dualities by integrating out some matter fields (see also \cite{WY} and \cite{Kapustin}). Recently, it was shown \cite{KWY3} that ${\cal N}=6$ dualities proposed by Aharony, Bergman and Jafferis \cite{ABJ} are descendant from Aharony dualities.  The fact that Aharony-type dualities generate a large class of dualities in three dimensions makes their verification and further understanding an important task.   

Such a verification was recently performed by Willett and Yaakov \cite{WY} who showed that partition functions on $S^3$ agree for theories which are related by Aharony duality.

In the present paper we verify that the superconformal indices of theories related by Aharony duality agree to a high order in the Taylor expansion for several low values of $N_c$ and $N_f$. This is of interest because agreement of indices is a check independent of agreement of partition functions on $S^3$.   We also discuss the role played in the duality by monopole operators. In particular, we discuss the matching of chiral rings in dual theories taking account of monopole operators.

\section{Index for ${\cal N}=2$ theories}

The superconformal index of an ${\cal N}=2$ superconformal theory on $S^2\times{\mathbb R}$ is defined by the expression
\begin{align}
{\cal I}(x, z_i)=Tr[(-1)^Fx^{E+j_3}\prod_{i}z_i^{F_i}]\label{ind}
\end{align}
where $F$ is the fermion number, $E$ is the energy, $j_3$ is the third component of spin and $F_i$ are charges of abelian flavor symmetries. As usual, contributions to the index come from states with $\{Q,Q^\dagger\}=E-r-j_3=0$ \cite{BBMR},\cite{Kim}. $r$ is the $R$-charge and $Q$ has spin $-1/2$.

An important feature of ${\cal N}=2$ superconformal theories in three dimensions is that the conformal dimensions of fields are not canonical in general and generically are irrational. The formula for the superconformal index of a theory with canonical conformal dimensions $\Delta_\Phi$ of chiral superfields $\Phi$ from the UV Lagrangian was obtained by Kim \cite{Kim} and recently generalized to any conformal dimensions by Imamura and Yokoyama \cite{IY}
\begin{align}
{\cal I}(x_2, z_i)=\sum_{\{n\}}\int[da]_{\{n\}}x^{E_0(\{n\})}e^{iS_{CS}^0(\{n\},a)}\prod_{i}z^{F_i^0}exp(\sum_{m=1}^\infty f(x^m, z_i^m, a^m))\label{index}
\end{align}
 The sum $\sum_{\{n\}}$ is over all GNO charges \cite{GNO} $\{n\}=(n_1,...,n_c)$ with $n_i\equiv w_i(H)$ where $w_i$ are the weights of the fundamental representation and $H$s are all element of a Cartan subalgebra defining a Dirac monopole. The integral whose measure depends on GNO charges is over a maximal torus of the gauge group, $E_0(\{n\})$ is the energy of a bare monopole with GNO charges $\{n\}$ and $F^0_i=-\sum_{\Phi}\sum_{\rho\in R_{\Phi}}|\rho(H/2)|F_{i\Phi}$ is its global charge under a global symmetry $U(1)_{i}$, the sum being over all gauge weights of all chiral fields with $F_{i\Phi}$ being their $U(1)_{F_i}$ charges. $S_{CS}^0(\{n\},a)$ is effectively the weight of the bare monopole with respect to the gauge group and $a$ is in a Cartan subalgebra.The function $f=f_{ch}+f_v$ depends on the content of vector multiplets and hypermultiplets. 
\begin{align}
& f_{ch}=\frac{1}{1-x^2}\sum_\Phi\sum_{\rho} x^{|\rho(H)|}(x^{\Delta_\Phi}e^{i\rho(a)}\prod_{i}z^{F_i}-x^{2-\Delta_\Phi}e^{-i\rho(a)}\prod_{i}z^{-F_i})\nonumber\\
& f_v=-\sum_{\alpha}x^{|\alpha(H)|}e^{i\alpha(a)}
\end{align}
The first sum in the expression for $f_{ch}$ is over all chiral multiplets $\Phi$. The second sum is over weights $\rho$ of the representations of the gauge group in which the chiral fields $\Phi$ live.  The contribution of the vector multiplet $f_v$ contains a sum over all roots $\alpha$ and does not contain any anomalous dimensions because it is assumed that the superconformal $R$-current at the IR fixed point is a linear combination of a UV $R$-current and some global  $U(1)$ symmetry current visible classically (in the UV). This guarantees that the vector multiplet retains its classical dimension. In general, the superconformal $R$-current can mix with accidental symmetry currents. In such a case the above formula for the index is not correct. We assume, following Gaiotto and Witten \cite{GW}, that this manifests itself in violation of unitarity bound on conformal dimensions of chiral operators including monopole operators, and thus, in principle we know when the formula for the index is correct.
The closed-form expression for the index is not known for nonabelian gauge theories\footnote{See paper \cite{KSV} for the abelian case.}, but a finite number of terms in its Taylor expansion around point $x=0$ can be computed on the computer. 

The fact that conformal dimensions $\Delta_\Phi$ are not known does not pose a problem if the goal is to perform a check of duality. As usual, the index can be computed as a path integral with (twisted) periodic boundary conditions along the time line ${\mathbb R}$. That is, it is a path integral on $S^2\times S^1$. There are many ways to put the theory on $S^2\times S^1$ parametrized by the choice of the $R$-current \cite{IY}. For the present theories any $R$-current is a linear combination $J_R=J_R^{UV}+\alpha J_{A}$ of the UV $R$-current $J_R$ and the global current $J_A$ generating the $U(1)_A$ symmetry.   For a special choice of the current, that is, for a special value of parameter $\alpha$ which determines anomalous dimensions of fields, the theory on $S^2\times S^1$ is superconformal.\footnote{This special value of the the parameter $\alpha=\Delta-1/2$ is determined by the extremization of the absolute value of the partition function of the theory put on $S^3$ with respect to $\Delta$ \cite{Jafferis}.} In this case the quantity computed by the path integral is the index in the sense of definition (\ref{ind}) with the trace over the Hilbert space of states living on $S^2$. For other values of the parameter it does not have this interpretation but it is nevertheless a quantity characterizing the theory which is independent of the description of the theory, that is, independent of a duality frame. Thus the 'indices' of dual theories must coincide as functions of the parameter $\alpha$. So we can introduce a new variable $y\equiv x^{\alpha}$ following \cite{IY} and compare the indices as functions of two variables $x$ and $y$.

\section{Aharony duality for unitary groups}

The duality relates two theories which we will call electric and magnetic. The electric theory is the ${\cal N}=2$ supersymmetric gauge theory with gauge group $U(N_c)$ with $N_f$ flavors of fundamental chiral fileds $Q_i$ and $N_f$ flavors of anti-fundamental chiral fields $\tilde Q^{\tilde\imath}$. The global symmetry group is $SU(N_f)\times\widetilde{SU(N_f)}\times U(1)_A\times U(1)_T\times U(1)_R$. The first two factors are flavor symmetries, the third factor is a rotation of both $Q_i$ and $\tilde Q^i$ by the same phase, $U(1)_R$ is the microscopic $R$-symmetry and $U(1)_T$ is the topological symmetry with the current $J^\mu=-\frac{1}{4\pi} \eps^{\mu\nu\rho}\Tr F_{\nu\rho}$ under which no elementary field is charged. We summarize the action of the global symmetry group in Table 1.

\begin{longtable}{|l|l|l|l|l|l|}
\hline
Fields & $U(1)_R$ & $U(1)_A$ & $SU(N_f)$ & $\widetilde{SU(N_f)}$ & $U(1)_T$\\
\hline
$Q$ & 1/2 & 1 & ${\bf N_f}$ & ${\bf 1}$ & 0\\
$\tilde Q$ & 1/2 & 1 & ${\bf 1}$ & ${\bf N_f}$ & 0\\
$M_i^{\tilde\jmath}$ & $1$ & $2$ & ${\bf N_f}$ & ${\bf N_f}$ & $0$\\
$v_{\pm}$ & $N_f-N_c+1$ & $-N_f$ & ${\bf 1}$ & ${\bf 1}$ & $\pm 1$\\
\hline
\caption{Global charges of fields of the electric theory.}
\end{longtable}

Here $M_i^{\tilde\jmath}\equiv Q_i\tilde Q^{\tilde\jmath}$ is the meson field and $v_{\pm}$ are monopole fields. In the ultraviolet theory the monopole operators are defined as disorder operators in the path integral \cite{BKW} with topological charges $\pm 1$. On the Coulomb branch below the Higgs scale with all charged fields integrated out they appear in the path integral as $\prod_{i=1}^{N_c}e^{\frac{\sigma_i\pm i\gamma_i}{N_c}}$ where $\sigma_i$ are real scalars from the vector multiplets of the broken gauge group $U(N_c)\to\prod_{i=1}^{N_c}U(1)_i$ and $\gamma_i$ are dualized photons. More precisely, in the UV description the correlation functions of monopole operators with fundamental fields are defined by performing the path integral over fields configuration having a Dirac monopole type singularity for gauge fields
\begin{align}
A^{N,S}=\frac{H}{2r}(\pm 1-\cos{\theta})d\phi
\end{align}
 together with the corresponding singularity $\sigma=-\frac{H}{2r}$ for the real scalar $\sigma$ in the vector multiplet at the insertion point to make the operator chiral. The GNO charges of monopole operators $v_{\pm}$ are $(\pm 1, \underbrace{0,..,0}_{N_c-1})$.

On the magnetic side is the ${\cal N}=2$ supersymmetric gauge theory with gauge group $U(N_f-N_c)$ with $N_f$ flavors of fundamental chiral fileds $q^i$ and $N_f$ flavors of anti-fundamental chiral fields $\tilde q_{\tilde\imath}$. In addition, there are two gauge-singlet chiral fields $v_\pm$  which correspond to the monopole operators of the electric theory and a gauge-singlet chiral field $M_i^{\tilde\jmath}$ which is a counterpart to the meson $Q_i\tilde Q^{\tilde\jmath}$. The theory has a  superpotential $W=M_i^{\tilde\jmath}q^iq_{\tilde\jmath}+v_+V_-+v_-V_+$ where $V_\pm$ are monopole chiral operators with GNO charges $(\pm 1,\underbrace{0,...,0}_{N_f-N_c-1})$. The representations of the fields under the action of the global symmetry group $SU(N_f)\times\widetilde{SU(N_f)}\times U(1)_A\times U(1)_T\times U(1)_R$ are written in Table 2.

\begin{longtable}{|l|l|l|l|l|l|}
\hline
Fields & $U(1)_R$ & $U(1)_A$ & $SU(N_f)$ & $\widetilde{SU(N_f)}$ & $U(1)_T$\\
\hline
$q$ & 1/2 & 1 & ${\bf N_f}$ & ${\bf 1}$ & 0\\
$\tilde q$ & 1/2 & 1 & ${\bf 1}$ & ${\bf N_f}$ & 0\\
$M_i^{\tilde\jmath}$ & $1$ & $2$ & ${\bf N_f}$ & ${\bf N_f}$ & 0\\
$v_{\pm}$ & $N_f-N_c+1$ & $-N_f$ & ${\bf 1}$ & ${\bf 1}$ & $\pm 1$\\
$V_{\pm}$ & $N_c-N_f+1$ & $N_f$ & ${\bf 1}$ & ${\bf 1}$ & $\pm 1$\\
\hline
\caption{Global charges of fields of the magnetic theory.}
\end{longtable}

Note that some of the 'elementary fields' -- $v_{\pm}$ are now charged under the topological symmetry. This is compatible with the invariance of the superpotential.
The only information about the superpotential in the formula for the index (\ref{index}) is the constraints on the superconformal IR $R$-charges of fields it provides.

We computed indices for several dual pairs of theories.

\subsection{Indices for dual pairs of theories with unitary gauge groups}

We use the notation $U(N_c)_{N_f}$ to denote the electric theory with gauge group $U(N_c)$ and $N_f$ pairs of fundamental and antifundamental chiral fieds. The magnetic theory with gauge group $U(N_c)$ and $N_f$ pairs of fundamentals and antifundamentals and additional singlets is denoted by $U(N_c)_{N_f}+M_i^{\tilde\jmath}+v_{\pm}$.

\begin{itemize}

\item[({\bf i})] Electric theory: $U(2)_2$. Magnetic theory: $U(0)+M_i^{\tilde j}+v_{\pm}$.

 In this case there is no vector mutiplet and no superpotential in the magnetic theory. The chiral fields $2\times 2$ matrix $M_i^{\tilde\jmath}$ and two $SU(2)_f\times\widetilde{SU(2)_f}$ flavor singlets $v_+$ and $v_-$ are free. The conformal dimension $\Delta\equiv\Delta(Q)=\Delta(\tilde Q)$ was computed in \cite{WY} to be $1/4$. This is one of the rare cases when the conformal dimension is rational. The conformal dimensions of the fields of the magnetic theory are easy to find using the duality dictionary. The conformal dimension $\Delta(M)=2\Delta$ of the $M_i^{\tilde\jmath}$ is twice the conformal dimension of $Q$ because these fields correspond to the meson of the electric theory. The conformal dimensions of singlet fields $v_{\pm}$ are equal to the conformal dimensions of bare monopole fields $(\pm 1,0)$ on the electric side: $\Delta(v_{\pm})=1/2$. Of course, this is obvious because all chiral fields of the magnetic theory are free and thus have conformal dimension one half.
To the second order in $x$ the index of the magentic theory is
\begin{align}
{\cal I}_B=1+6x^{1/2}+21x+50x^{3/2}+90x^2+{\cal O}(x^{5/2})
\end{align}  
The first term is the contribution of the vacuum and the second term comes from the six free chiral fileds. The contribution to the index on the $A$-side comes from sectors with different GNO charges. It is summarized in Table 3.
\begin{longtable}{|l|l|}
\hline
GNO charges & Index contribution\\
\hline
$(0,0)$ & $1+4x^{1/2}+10x+20x^{3/2}+27x^2$\\
$(1,-1)$ & $x$\\
$(2,-2)$ & $x^2$\\
\hline
$(1,0)$ & $x^{1/2}+4x+9x^{3/2}+16x^2$\\
$(2,-1)$ & $x^{3/2}$\\
\hline
$(2,0)$ & $x+4x^{3/2}+9x^2$\\
$(3,-1)$ & $x^2$\\
\hline
$(3,0)$ & $x^{3/2}+4x^2$\\
\hline
$(4,0)$ & $x^2$\\
\hline
\caption{Contribution to the index from different GNO sectors in $U(2)_2$ theory.}
\end{longtable}
Summation of these contributions over the topological charges (the contribution from the negative topological charges are the same as from the positive ones) reproduces the answer on the magnetic side, which consitutes a nontrivial check of the duality.

\end{itemize}

In general, we do not expect the GNO charges within a sector with a fixed $U(1)_T$ charge to mark sectors in the Hilbert space of the theory because they do not arise from any conserved currents. Rather, it is an artifact of the weakly coupled description of the theory. We saw it in the previous paper \cite{BK2} where the indices of dual theories were in agreements within a given topological sector only after summation over all GNO charges and there was no mapping of GNO charges between dual theories. This was also noticed in \cite{JY}.

However, in certain situations GNO charges may acquire invariant meaning if they correlate with other quantum numbers. This is the present case. For each value of the $U(1)_T$ charge and the $U(1)_A$ charge the GNO charge of a bare monopole is determined uniquely. We list the global charges of some of the low-energy bare monopoles in Table 4.
\begin{longtable}{|l|l|l|l|}
\hline
Bare monopole & Conformal dimension & Topological charge & $U(1)_A$-charge\\
\hline
$(1,0)$ & 1/2 & 1 & -2\\
$(-1,0)$ & 1/2 & -1 & -2\\
$(1,-1)$ & 1 & 0 & -4\\
$(2,-2)$ & 2 & 0 & -8\\
$(2,-1)$ & 3/2 & 1 & -6\\
$(1,1)$ & 3 & 2 & -4\\
$(2,0)$ & 1 & 2 & -4\\
$(3,-1)$ & 2 & 2 & -8\\
$(3,0)$ & 3/2 & 3 & -6\\
$(4,0)$ & 2 & 4 & -8\\
\hline
\caption{Quantum numbers of bare monopole operators in $U(2)_2$ theory.}
\end{longtable}
The duality relates monopole operators of the electric theory to (composite) chiral fields of the magnetic theory. Using matching of quantum numbers it is easy to establish a dictionary for this correspondance. For some of the low-dimension operators it is
\begin{longtable}{|l|}
\hline
Chiral operator and OPE\\
\hline
$v_+\equiv T_{(1,0)}$\\
$v_-\equiv T_{(-1,0)}$\\
$T_{(1,-1)}\sim v_+v_-$\\
$T_{(2,-2)}\sim v_+^2v_-^2$\\
$T_{(2,-1)}\sim v_+^2v_-$\\
$T_{(1,1)}\sim M^2v_+^3v_-$\\
$T_{(2,0)}\sim v_+^2$\\
$T_{(3,-1)}\sim v_+^3v_-$\\
$T_{(3,0)}\sim v_+^3$\\
$T_{(4,0)}\sim v_+^4$\\
\hline
\caption{Mapping of chiral operators under duality. $M^2$ is the $SU(2)\times\widetilde{SU(2)}$ flavor singlet quadratic in meson fields.}
\end{longtable}

\begin{itemize}

\item[({\bf ii})] Electric theory is $U(2)_3$, magnetic theory is $U(1)_3+M_i^{\tilde\jmath}+v_{+}+v_-$.

In this case the conformal dimensions of all fields are irrational and $\Delta\equiv\Delta(Q)=\Delta(\tilde Q)\approx 0.3417$\footnote{We took the approximate values of conformal dimensions from \cite{WY}.}. We introduce additional variable $y\equiv x^{2\Delta-1}$ and expand the indices of both theories in powers of $x$. The contribution from different topological and GNO sectors are given in Tables 9 and 10 in Appendix A. We find a perfect agreement for each value of the topological charge up to the third power in $x$.

\item[({\bf iii})] Electric theory: $U(2)_4$, magnetic theory: $U(2)_4+M_i^{\tilde\jmath}+v_{\pm}$.
The conformal dimension of $Q$ is $\Delta\approx 0.3852$.
Naively, the magnetic theory contains more degrees of freedom than the electric theory by weak-coupling counting. Nevertheless, they flow to the same infrared fixed point. The indices agree in each topological sectors of both theories up to at least the third power in $x$ (Tables 11 and 12 in Appendix A).

\item[({\bf iv})]

As our last check of the duality for unitary groups we chose the following pair. Electric theory is $U(3)_4$, and magnetic theory is $U(1)_4+M_i^{\tilde\jmath}+v_{+}+v_-$ ($\Delta\approx 0.3058$).
We found agreement of indices for each topological sector up to the fourth power in $x$ (Tables 13 and 14).

\end{itemize}

\section{Chiral ring}

\subsection{Examples}

\begin{itemize}

\item[{\bf (i)}]There are two ways to look at the Table 5. One way is to view it as a correspondence between operators on different sides of duality. Another way is a relation in the chiral ring of the electric theory if we regard $v_{\pm}$ as chiral monopole operators  with GNO charges $(\pm 1,0)$.  In particular, we see that the chiral ring is generated by 6 generators -- chiral fields $M_i^{\tilde j}$ and two chiral monopole operators $v_\pm$.

\item[{\bf (ii)}]The situation is more involved for greater number of flavors and larger gauge groups. As the next simplest case we consider the chiral rings of the dual pair of theories: the electric theory $U(2)_3$ and the magnetic theory $U(1)_3+M_i^{\tilde\jmath}+v_{+}+v_-$. 

\end{itemize}

First we look at the magnetic side. The generators of the chiral ring include eleven operators: mesons $M_i^{\tilde\jmath}\equiv Q_i\tilde Q^{\tilde\jmath}$ and $v_{\pm}$. Other candidates for generators are monopole operators. The monopole operators $V_{\pm}$ having GNO charges $(\pm 1)$ are dismissed right away because they are $Q$-exact due to the presence of the superpotential $v_+V_-+v_-V_+$. There remain monopole operators with higher values of GNO charge. However, they are also $Q$-exact because they are just powers of $V_{\pm}$. Namely, $V_{n>0}=V_+^n$ and $V_{n<0}=V_-^{-n}$. This can be seen from the fact that all global charges agree and contribution of these operators to the index cancels. This does not consitute a proof.  Nevertheless, it appears to be very natural. Thus we assume that the eleven chiral operators are all generators of the chiral ring. We provide an additional argument in favor of this conclusion later.

On the electric side of duality there are chiral operators: $M_i^{\tilde\jmath}$ and $v_{\pm}$ where the last two are now monopole operators. We should address the question of whether some of the monopole operators are in fact not generated by $M_i^{\tilde\jmath}$ and $v_{\pm}$. For example, are there any monopole operators whose quantum numbers are such that no monomial in the generators $M_i^{\tilde\jmath}$ and $v_{\pm}$ can reproduce them? Naively, such a monopole operator does exist. In fact, there are many of them and they all are generated in terms of quantum numbers by the operator corresponding to the bare monopole state $\ket{1,1}$ which has GNO charge $(1,1)$. To understand the origin of this phenomenon one should recall the framework in which the monopole operators are treated. We will discuss the general case $U(N_c)_{N_f}$ and use the duality conjecture to recover some information about monopole operators in the next few paragraphs and then return to the special case $U(2)_3$ to illustrtate the general conclusions that we make. 

\subsection{General discussion}

The definition of monopole operators as a certain class of disordered operators is only constructive in weakly interacting theories. When the theory of interest is not weakly coupled yet supersymmetric one can proceed in two steps to make use of these operators. First, the theory is put on $S^2\times R$, that is, radially quantized. Second, a supersymmetic deformation to a weak coupling is performed. In the first step monopole operators become states in the radially quantized picture as all local operators do. In the second step the supersymmetry guarantees that some information about the original theory is preserved in the deformed theory which describes dynamics of free fields quanta in the classical monopole backgrounds parametrized by GNO charges. The Fock vacua in every GNO sector of this theory are the bare monopoles. The index formula (\ref{index}) computes the index (\ref{ind}) of this free theory which by the supersymmetry of the deformation is the index of the original radially quantized theory. This is an example of the preserved information. Another example is the spectrum of chiral scalars which are bottom components of different current multiplets \cite{BK,BK2}.

Unfortunately, the chiral ring as a vector space is not part of the structure of the original theory preserved by the deformation. We show it in the next subsection. Two things can happen. First, a state corresponding to a nontrivial element of the chiral ring of the original theory may become $Q$-exact when the deformation is switched on if there are states with appropriate quantum numbers to pair up with it. Then the energy of this long multiplet may be changed in the deformed theory so that no traces of the original state are seen in the deformed theory. Even if the energy is not changed we do not pay attention to long multiplets in the deformed theory because they will remain long when the deformation is switched on and what happens to them is anyone's guess. The $U(2)_2$ theory provides an example -- in the deformed electric theory there is no state corresponding to chiral operator $v_{+}v_{-}M_i^{\tilde\jmath}$. There is a manifestation of this in the index -- there is no contribution with the quantum numbers of $v_{+}v_{-}M_i^{\tilde\jmath}$ (see Tables 3 and 4). In the magnetic theory this happens because the contribution of $v_{+}v_{-}M_i^{\tilde\jmath}$ is canceled by the contribution of the BPS spinor $\Psi^\dagger{}_i^{\tilde\jmath}$, the conjugate of the superpartner of $M_i^{\tilde\jmath}$.  Second, there may appear accidental $Q$-cohomology classes in the deformed theory by essentially the opposite process. In fact, as explained below, these two processes become more likely with the increase of the energy of states and rank of the gauge group.

Yet, some low energy states are in fact protected. These are states corresponding to operators $M^i_{\tilde\jmath}$ and $v_{\pm}$ that are naturally expected to be the complete set of generators of the chiral ring. Of course, the presence of meson operators in the chiral ring of the electric theory is obvious, and, due to the duality, the presence of monopole operators $v_{\pm}$ is guaranteed. From the point of view of the electric theory their presence is ensured as they are the lowest energy states in the sector with topological charge one and they cannot pair up with fermions of higher energy. More precisely, for a BPS scalar to become a part of a longer multiplet there must be a fermion available with appropriate quantum numbers. In particular, by unitarity, its energy must be less than that of the scalar.

There remains a possibility that some other monopole operators can complete the set of generators of the chiral ring. Below we argue that the assumption that this does not happen is consistent with the information preserved along the deformation. 

\subsection{Scalar BPS states in the deformed theory}

The Hilbert space of the deformed theory is the direct sum of Fock spaces whose vacua are bare monopole states with different GNO charges. All these vacua are BPS scalars. Other BPS scalar states are obtained by acting on the bare monopoles with the creation operators corresponding to the fields of the theory. It is not a problem to obtain scalar states in this way but the BPS condition is quite restrictive. Consider a bare monopole state $\ket{n_1,...,n_{N_c}}$ with GNO charges $(n_1,...,n_{N_c})$. A matter field creation operator $\varphi_{i}$ with gauge index $i$ interacts with $n_i$ units of magnetic charge. As a result \cite{BKW,BKW2} it obtaines 'anomalous' spin with minimal value $j_0=\frac{|n_i|}{2}$ if $\varphi$ was a scalar field and $j_0=\frac{|n_i|-1}{2}$ if it was a spinor.  The energy of this mode is also changed compared to the case when the mode does not interact with magnetic flux. We list the different modes and their energies when they are coupled to $n$ units of magnetix flux in Table 6 below.

\begin{longtable}{|c|c|c|c|}
  \hline{\rm fields}&$U(1)_R*$&Spin&Energy\\
  \hline 
$Q^\dagger$&$\frac{1+\alpha}{2}$&$j_0=\frac{|n|}{2}$&$j+\frac{1+\alpha}{2}=r+j$\\
$\tilde Q^{\dagger}$&$\frac{1+\alpha}{2}$&$j_0=\frac{|n|}{2}$&$j+\frac{1+\alpha}{2}=r+j$\\
$\psi_{Q}$ & $\frac{1-\alpha}{2}$&$j_0=\frac{|n|+1}{2}$&$j+\frac{1-\alpha}{2}=r+j$\\
$\psi_{\tilde Q}$&$\frac{1-\alpha}{2}$&$j_0=\frac{|n|+1}{2}$&$j+\frac{1-\alpha}{2}=r+j$\\ 
$\psi_{Q}^\dagger$ & $-\frac{1-\alpha}{2}$ & $j_0=\frac{|n|-1}{2}$ & $j+\frac{1+\alpha}{2}>r+j$\\
$\psi_{\tilde Q}^\dagger$ & $-\frac{1-\alpha}{2}$ & $j_0=\frac{|n|-1}{2}$ & $j+\frac{1+\alpha}{2}>r+j$\\
  \hline 
 $a^{(1)}$ & 0 & $j_0=\frac{|n|}{2}+1$ & $j=r+j$\\
 $a^{(2)}$ & 0 & $j_0=\frac{|n|}{2}+1$ & $j+1>r+j$\\
 $\lambda$ &-1& $j_0=\frac{|n|+1}{2}$ & $j>r+j$\\
 $\lambda^\dagger$ &1& $j_0=\frac{|n|-1}{2}$ & $j+1=r+j$\\
  \hline
 \caption{Quantum numbers of fields and supercharges.}
\end{longtable}

  Here $U(1)_R*$ is the IR superconformal $R$-symmetry and $r$ is its charge. The last four modes come from the vector mutiplet. A scalar state is BPS iff its quantum numbers satisfy the relation $E=r$. This requirement can be met only if the modes that excite the bare monopole are modes of scalar fields $Q_i$ or $\tilde Q^{\tilde\jmath}$ that do not interact with magnetic flux or modes of gluino $\lambda^\dagger_{ij}$ that interact with one unit of the magnetic flux $|n_i-n_j|=1$. Here $i$ and $j$ are gauge indices. In the first case this means that among GNO charges $(n_1,...,n_{N_c})$ at least one must be zero $n_i=0$. In the second case the difference of at least two GNO charges must be one. Moreover, one must use at least two gaugino modes to guarantee gauge invariance. For example, the gauge invariant state built on $\ket{2,1}$ is $\lambda^\dagger_{12}\lambda^\dagger_{21}\ket{2,1}$.

 So, a scalar BPS state in the deformed theory is either a bare monopole with arbitrary GNO charges or a bare monopole excited with free squark modes and/or gluino modes interacting with one unit of magnetic flux.  All gauge indices of the squark and gluino modes must be contracted in a gauge invariant way. Here gauge invariance is with respect to the unbroken by the fluxes subgroup $U(N_1)\times\dots\times U(N_k)\subset U(N_c)$. Note that the number of squark modes $Q$ must be equal the number of squark modes $\tilde Q$ for the state to be gauge invariant.

Now we can look for counterparts of the chiral ring operators in the deformed theory. For a monomial in the monopole operators $v_{\pm}$, comparison of quantum numbers gives $v_+^nv_-^m\to\ket{n,-m,\underbrace{0,...,0}_{N_c-2}}$ where the ket-vector is the bare monopole state with GNO charges $(n,-m,\underbrace{0,...,0}_{N_c-2})$. Multiplying this operator by a meson field $M_i^{\tilde\jmath}$ naturally corresponds to $Q_i\tilde Q^{\tilde\jmath}\ket{n,-m,\underbrace{0,...,0}_{N_c-2}}$ where the gauge indices of scalar modes of squarks run over $N_c-2$ values corresponding to the unbroken gauge group $U(N_c-2)\subset U(N_c)$ and are contracted properly to form a gauge singlet. There must be no gauge indices corresponding to the commutant of the $U(N_c-2)$ in $U(N_c)$ because such modes interact with the monopole background and as a result their energy is increased  \cite{BKW}, \cite{BKW2} which makes it impossible to build BPS scalars with them. Multiplying by more powers of mesons corresponds to putting more squarks modes on the bare monopole. If $N_c\le 2$, then there is no state in the deformed theory corresponding to the operator $v_+^nv_-^mM_i^{\tilde\jmath}$.

  The next question is whether there is a scalar BPS monopole operator with such quantum numbers that it cannot be generated by mesons $M_j^{\tilde\jmath}$ and monopole operators $v_{\pm}$, which then is a new generator of the chiral ring. As usual, the direct analysis of the original theory which is strongly coupled is out of reach, so one can try looking at the deformed dual theories.  

If in the deformed electric theory there is a BPS state with quantum numbers which cannot be reproduced by a monomial in $M^i_{\tilde\jmath}$ and $v_{\pm}$, then, by the above argument, this state must be a bare monopole excited with free squark modes and/or gluino modes. The free squark modes correspond to (a product of) meson operators, so we can strip the state of them. This new state corresponds to a BPS monopole operator which is still not generated by the mesons and monopole operators $v_{\pm}$. Now we make use of the conjectured duality.  In the dual magnetic theory this operator correspond to a (dressed) monopole state. If it contains free squark modes, we repeat the procedure again to obtain a monopole state which is either bare or excited with only gluino modes. Then we again look at the electric side, and so on. This process reduces energy, so it must stop at some step. It stops only if a monopole operator not generated from the mesons and $v_{\pm}$ corresponds to states in the deformed theories which are both either bare monopoles or bare monopoles excited with gluino modes.  However, bare monopoles or bare monopoles excited with only gluino modes on the different sides of the duality can never have the same $U(1)_A$ charge. Indeed, the $U(1)_A$ charges of such monopoles on the electric side $A=-N_f\sum_{i=1}^{N_c}|n_i|$ are always negative while the charges of monopoles on the  magnetic side $A=N_f\sum_{j=1}^{N-f-N_c-1}|n_j|$ are always positive. Thus, they never match. So the assumption that mesons and minimal monopole operators $v_{\pm}$ exhaust the generators of the chiral ring is consistent with the information preserved by the deformation. Moreover, the chiral ring is freely generated by them as long the IR superconformal R-current is not accidental. On the magnetic side this is obvious in view of the absence of a superpotential monomial depending on $v_{\pm}$ and $M_i^{\tilde j}$ simultaneously. On the electric side this can be proved not using the duality conjecture -- matching quantum numbers of any relation between them lead to negative energies of either the mesons or $v_{\pm}$\footnote{See Appendix B}.

The conclusion is that in all Aharony-type theories  with arbitrary $N_f$ and $N_c$\footnote{As long as $N_f$ is big enough compared to $N_c$ for an accidental $R$-charge not to appear.} the deformation does not preserve the chiral ring as a vector space. Indeed, if the chiral ring is generated not by only mesons and minimal monopole operators $v_{\pm}$ then by the above reasoning there cannot be one-to-one correspondance between BPS scalar states in the deformed theories and chiral operators in the original one.  If, on the other hand, the entire chiral ring is generated by mesons and $v_{\pm}$, then it is not preserved by the deformations either, because there are many BPS monopoles in the deformed theories (for $N_c>2$) whose quantum numbers forbid them to correspond to monomials in mesons and $v_{\pm}$. Thus the spectra of scalar BPS states in the original and deformed theories are not the same.   

\subsection{Illustration of the general conclusions}

Returning to the state $\ket{1,1}$ in the $U(2)_3$ theory, the most natural explanation of its appearence in view of the duality is that it is accidental in the deformed theory. When the interactions are switched on it gets paired up with a fermion and is not present in the original theory as a nontrivial element of the chiral ring. In other words, it is zero in the chiral ring. Indeed, its contribution to the index is $x^3y^{-3}$ while there are fermionic monopole operators with GNO charge $(2,0)$ which contribute $-18x^3y^{-3}$ to the index. Among these fermionic operators there are those with quantum numbers necessary for $\ket{1,1}$ to become their superdescendant once the interactions are turned back on. 

This example illustrates a general fact about monopole operators in the ${\cal N}=2$ SQCD theories.  Assumption of the completeness of the chiral ring freely generated by meson operators $M$ together with minimal monopole operators $v_{\pm}$  leads to the conclusion that all bare monopole states in the deformed theory with GNO charges different from $(n,-m,0,...,0)$ are accidental BPS states. We provide evidence in favor of this statement in Appendix C. 

As an additional example we consider the bare monopole state $\ket{1,1,0}$ in $U(3)_4$ theory from example ${\bf (iv)}$. Its contribution to the index is $x^8y^{-4}z^{-8}$ where the power of $z$ indicates the $U(1)_A$ charge. In the same topological sector monopole states with GNO charge $(2,0,0)$ contribute $x^4y^{-4}z^{-8}-32x^8y^{-4}z^{-8}+{\cal O}(x^{10})$. The 32 fermionic states have the form $\bar\psi^i Q_j\ket{2,0,0}$ and $\bar{\tilde\psi}_{\tilde\imath}\tilde Q^{\tilde\jmath}\ket{2,0,0}$ where $(i,j)$ are indices of flavor group $SU(4)$, $(\tilde i,\tilde j)$ are indices of flavor group $\widetilde{SU(4)}$ and gauge indices corresponding to the unbroken $U(2)$ are contracted properly and not shown. In terms of the representation of the flavor group $SU(4)\times\widetilde{SU(4)}$ the 32 fermions are $(\bar{\bf 4},{\bf 1})\times({\bf 4},{\bf 1})+({\bf 1},\bar{\bf 4})\times({\bf 1},{\bf 4})$. There are two flavor singlets among them, one of which can pair up with the bare monopole $\ket{1,1,0}$. 

 Another conclusion is that not all elements of the chiral ring are present in the deformed theory. For instance, in the example ${\bf (i)}$ there is no state in the deformed theory corresponding to operator $v_+v_-M^i_{\tilde\jmath}$. This is possible because this state does not make a distinguished contribution to the index. Indeed, the term $20x^{3/2}$ originates from states with $U(1)_A$ charge $6$ instead of $-2$ which would be if contribution of the operator $v_+v_-M^i_{\tilde\jmath}$ was not canceled by a potential fermionic superpartner.\footnote{This is seen when the additional parameter $z$ corresponding to the $U(1)_A$ symmetry introduced into the index.}

Finally, an important conclusion is that GNO charges do not parametrize sectors in the Hilbert space as charges of global symmetries do. They are just labels of  operators or states in the radially quantized picture. This follows from the fact that the ${\cal N}=2$ $U(N)$ SQCD with $N_f$ flavors of quarks and $N_f$ flavors of antiquarks contain bare monopole operators that must be\footnote{Assuming validity of Aharony duality which is now well tested.} superdescendants of fermions that have different GNO charges, because these fermions have lower conformal dimension and the bare monopoles are the lowest conformal dimension operators with given GNO charges. The same conclusion can be reached if one notes that all monopole operators on the $B$-side are superdescendants, so the monopole operators of the electric theory which are nontrivial elements of the chiral ring do not correspond to any monopole operators of the magnetic theory. Rather, they correspond to operators which are generated by elementary fields.

\section{Aharony duality for symplectic groups}

The duality for symplectic groups is quite similar to the case of unitary groups. The electric theory is a $USp(2N_c)$ ${\cal N}=2$ gauge theory with $2N_f$ chiral multiplets in the fundamental representation. The composite chiral gauge invariant fields include the meson $M_{ij}\equiv Q_iQ_j$ and the monopole field $Y$. Their quantum numbers are displayed in Table 7.
\begin{longtable}{|l|l|l|l|}
\hline
Fields & $U(1)_R$ & $U(1)_A$ & $SU(2N_f)$ \\
\hline
$Q$ & 1/2 & 1 & ${\bf 2N_f}$ \\
$M$ & $1$ & $2$ & ${\bf N_f(2N_f-1)}$\\
$Y$ & $2(N_f-N_c)$ & $-2N_f$ & ${\bf 1}$\\
\hline
\caption{Global charges of fields of the electric theory.}
\end{longtable}
The dual theory is an $USp(2(N_f-N_c-1))$ ${\cal N}=2$ gauge theory with $2N_f$ chiral multiplets $q_i$ in the fundamental representation together with singlet chiral fields $M_{ij}$ and $Y$ which correspond to the composite chiral fields on the electric theory. There is a superpotential $W=M_{ij}q_{i}q_{j}+Y\tilde Y$ where $\tilde Y$ is the monopole field in the magnetic theory. The global charges of all fields are written in Table 8
\begin{longtable}{|l|l|l|l|}
\hline
Fields & $U(1)_R$ & $U(1)_A$ & $SU(2N_f)$\\
\hline
$q$ & 1/2 & 1 & ${\bf 2N_f}$\\
$M$ & $1$ & $2$ & ${\bf N_f(2N_f-1)}$\\
$Y$ & $2(N_f-N_c)$ & $-2N_f$ & ${\bf 1}$\\
$\tilde Y$ & $-2(N_f-N_c-1)$ & $2N_f$ & ${\bf 1}$\\ 
\hline
\caption{Global charges of fields of the magnetic theory.}
\end{longtable}
Unlike the previously discussed theories with unitary gauge groups, the gauge groups in the present case are simple which means there is no topological current. The monopole operator $Y$ does not carry any quantum numbers in addition to the perturbative ones\footnote{By perturbative quantum numbers we mean Noether charges associated with symmetries of the UV Lagrangian.}. The GNO charges are merely labels distinguishing different operators.  When comparing the indices of dual theories we must sum over GNO charges.   The GNO charges of $Y$ are $(1,\underbrace{0,...,0}_{N_c-1})$ and those of $\tilde Y$ are  $(1,\underbrace{0,...,0}_{N_f-N_c-2})$. 
We compared the indices for the following three dual pairs of theories and found complete agreement in the lower orders in $x$. Similarly to the case of unitary gauge groups the subscript of the gauge group stands for $N_f$.

\begin{itemize}

\item[{\bf(i)}] Electric theory: $USp(2)_3$, magnetic theory: $USp(2)_3+M+Y$.
The index is
\begin{align}
{\cal I}=1+15xy-36x^2+105x^2y^2+x^2y^{-6}+21x^3y^{-1}-384x^3y+490x^3y^3+x^3y^{-9}
\end{align}
where $y\equiv x^{2\Delta-1}$. The contributions from different GNO sectors are summarized in Tables 14 and 15 in Appendix A.

\item[{\bf(ii)}] Electric theory is $USp(4)_5$, magnetic theory is $USp(4)_5+M+Y$.

The index is
\begin{align}
& {\cal I}=1+45xy-100x^2+xy^{-5}+1035x^2y^2+45x^2y^{-4}+x^2y^{-10}+55x^3y^{-1}-\nonumber\\
& 4400x^3y+16005x^3y^3-99x^3y^{-5}+825x^3y^{-3}+45x^3y^{-9}+x^3y^{-10}
\end{align}
The contribution from different GNO sectors are written down in Tables 15 and 16 in Appendix A.

\end{itemize}

Arguments analogous to those for unitary gauge groups make plausible the assumption that the chiral rings of symplectic theories of Aharony types are freely generated by meson operators $M$ and operators $Y$ which are monopole operators of minimal GNO charges for electric theories and fundamental fields for magnetic theories. Analogously to the case of unitary gauge groups all bare monopoles with GNO charges different from $(n,0,...,0)$ are accidental BPS states in the deformed theory. The states $\ket{n,0,...,0}$ correspond to the chiral operator $Y^n$.

For instance, in the second example ${\bf (ii)}$ some bare monopole states on the electrical side are not generated  by only $Y$ and the mesons. There is $\ket{1,1}$ among the states, whose energy makes it impossible for it to correspond to any monomial in $Y$ and $M$s. Therefore, one expects that it gets paired up with a fermion on the way from the weak coupling to the original theory and is not present as a nontrivial element of the chiral ring in the original theory. Indeed, there is an indication of that in the index. The contribution of $\ket{1,1}$ is $x^4y^{-10}$ is canceled by the contribution $-100x^4y^{-10}$ of fermionic excited monopole with GNO charge $(2,0)$. In other words, the index suggests that it pairs up with a fermion with GNO charge $(2,0)$ and appropriate $U(1)_A$ charge.

 In the example ${\bf (i)}$ all bare monopole operators of the electric theory are nontrivial elements of the chiral ring and generated by the minimal bare monopole operator $Y$: if $T_{n>0}$ is the bare monopole operator with GNO charge $n>0$, then $T_{n>0}=Y^n$.\footnote{The Weyl group of $USp(2)=SU(2)$ identifies GNO charges $n$ and $-n$, and we choose representatives $n>0$.}

\section{Acknoweldgements}

I would like to thank Anton Kapustin for reading the draft of the paper and useful suggestions.

\section{Appendix A. Contribution to indices from different GNO sectors}

\begin{longtable}{|l|l|}
\hline
GNO charges & Index contribution\\
\hline
$(0,0)$ & $1+9xy-18x^2+45x^2y^2+9x^3y^{-1}-144x^3y+164x^3y^3$\\
$(1,-1)$ & $xy^{-3}$\\
$(2,-2)$ & $x^2y^{-6}$\\
$(3,-3)$ & $x^3y^{-9}$\\
\hline
$(1,0)$ & $x^{1/2}y^{-3/2}+9x^{3/2}y^{-1/2}-17x^{5/2}y^{-3/2}+36x^{5/2}y^{1/2}$\\
$(2,-1)$ & $x^{3/2}y^{-9/2}$\\
$(3,-2)$ & $x^{5/2}y^{-15/2}$\\
\hline
$(1,1)$ & $x^3y^{-3}$\\
$(2,0)$ & $xy^{-3}+9x^2y^{-2}-18x^3y^{-3}+36x^3y^{-1}$\\
$(3,-1)$ & $x^2y^{-6}$\\
$(4,-2)$ & $x^3y^{-9}$\\
\hline
$(3,0)$ & $x^{3/2}y^{-9/2}+9x^{5/2}y^{-7/2}$\\
$(4,-1)$ & $x^{5/2}y^{-15/2}$\\
\hline
$(4,0)$ & $x^2y^{-6}+9x^3y^{-5}$\\
$(5,-1)$ & $x^3y^{-9}$\\
\hline
$(5,0)$ & $x^{5/2}y^{-15/2}$\\
\hline
$(6,0)$ & $x^3y^{-9}$\\
\hline
\caption{Contribution to the index from different GNO sectors in $U(2)_3$ theory.}
\end{longtable}

\begin{longtable}{|l|l|l|}
\hline
GNO charge & Top. charge & Index contribution\\
\hline
$0$ & 0 & $1+xy^{-3}+9xy+x^2y^{-6}-20x^2+45x^2y^2+x^3y^{-9}-2x^3y^{-3}+$\\
 & & $27x^3y^{-1}-162x^3y^{-1}+166x^3y^3$\\
\hline
 & 1 & $x^{1/2}y^{-3/2}+x^{3/2}y^{-9/2}+9x^{3/2}y^{-1/2}-x^{3/2}y^{3/2}+x^{5/2}y^{-15/2}-$\\
& & $19x^{5/2}y^{-3/2}+45x^{5/2}y^{1/2}-9x^{5/2}y^{5/2}$\\
\hline 
& 2 & $xy^{-3}+x^2y^{-6}+9x^2y^{-2}-x^2+x^3y^{-9}-19x^3y^{-3}+45x^3y^{-1}-9x^3y^{1}$\\
\hline
 & 3 & $x^{3/2}y^{-9/2}+x^{5/2}y^{-15/2}+9x^{5/2}y^{-7/2}-x^{5/2}y^{-3/2}$\\
\hline
 & 4 & $x^2y^{-6}+x^3y^{-9}+9x^3y^{-5}-x^3y^{-3}$\\
 & 5 & $x^{5/2}y^{-15/2}$\\
\hline
& 6 & $x^3y^{-9}$\\
\hline
$1$ & 0 & $x^2+x^3y^{-3}-9x^3y^{-1}+9x^3y-x^3y^3$\\
 & 1 & $x^{5/2}y^{-3/2}$\\
 & 2 & $x^3y^{-3}$\\
\hline
$-1$ & 0 & $x^2+x^3y^{-3}-9x^3y^{-1}+9x^3y-x^3y^3$\\
 & 1 & $x^{3/2}y^{3/2}+x^{5/2}y^{-3/2}-9x^{5/2}y^{1/2}+9x^{5/2}y^{5/2}$\\
 & 2 & $x^2+x^3y^{-3}-9x^3y^{-1}+9x^3y-x^3y^3$\\
 & 3 & $x^{5/2}y^{-3/2}$\\
 & 4 & $x^3y^{-3}$\\
\hline
$-2$ & 2 & $x^3y^3$\\
\hline
\caption{Contribution to the index from different GNO sectors in $U(1)_3+M+v_{\pm}$ theory. GNO charge coincides with the topological charge for the bare monopole, but different for excited states due to the fact that fields $v_{\pm}$ carry topological charge.}
\end{longtable}

\begin{longtable}{|l|l|}
\hline
GNO charges & Index contribution\\
\hline
$(0,0)$ & $1+16xy-32x^2+136x^2y^2+16x^3y^{-1}-480x^3y+800x^3y^3$\\
$(1,-1)$ & $x^2y^{-4}$\\
$(2,-2)$ & $x^4y^{-8}$\\
\hline
$(1,0)$ & $xy^{-2}+16x^{2}y^{-1}-31x^{3}y^{-2}+100x^{3}$\\
$(2,-1)$ & $x^{3}y^{-6}$\\
\hline
$(2,0)$ & $x^2y^{-4}+16x^3y^{-3}$\\
\hline
$(3,0)$ & $x^{3}y^{-6}$\\
\hline
\caption{Contribution to the index from different GNO sectors in $U(2)_4$ theory.}
\end{longtable}

\begin{longtable}{|l|l|l|}
\hline
GNO charge & Top. charge & Index contribution\\
\hline
$(0,0)$ & 0 & $1+16xy+x^2y^{-4}-34x^2+136x^2y^2+x^4y^{-2}+16x^3y^{-1}+$\\
 & & $-512x^3y+816x^3y^3+16x^3y^5$\\
\hline
 & 1 & $xy^{-2}-xy^{2}+16x^{2}y^{-1}-16x^{2}y^{3}+x^{3}y^{-6}-$\\
& & $33x^{3}y^{-2}+136x^3+32x^{3}y^{2}-136x^{3}y^{4}$\\
\hline 
& 2 & $x^2y^{-4}-x^2+16x^3y^{-3}-16x^3y^{2}$\\
\hline
 & 3 & $x^{3}y^{-12}-x^{3}y^{-2}$\\
\hline
$(1,-1)$ & 0 & $x^2y^4-16x^3y^{3}+16x^3y^5$\\
 & 1 & $x^{3}y^{2}-x^3y^6$\\
 \hline
$(1,0)$ & 0 & $x^2-x^2y^{4}+16x^3y-16x^3y^5$\\
 & 1 & $x^{3}y^{-2}-x^{3}y^{2}$\\
 \hline
$(0,-1)$ & 0 & $x^2-x^2y^{4}+16x^3y-16x^3y^5$\\
 & 1 & $xy^2+16x^2y^3-36x^3+x^3y^{-2}-33x^3y^2+136x^3y^4+x^3y^6$\\
 & 2 & $x^2-x^2y^{4}+16x^3y-16x^3y^5$\\
 & 3 & $x^{3}y^{-2}-x^3y^2$\\
 \hline
$(0,-2)$ & 1 & $x^3y^{2}-x^3y^6$\\
 & 2 & $x^2y^4+16x^3y^5$\\
 & 3 & $x^3y^{2}-x^3y^6$\\
  \hline
$(0,-3)$ & 3 & $x^3y^6$\\
\hline
$(1,-2)$ & 1 & $x^3y^6$\\
\hline
\caption{Contribution to the index from different GNO sectors in $U(2)_4+M+v_{\pm}$ theory.}
\end{longtable}

\begin{longtable}{|l|l|}
\hline
GNO charges & Index contribution\\
\hline
$(0,0,0)$ & $1+16xy+136x^2y^2+816x^3y^3-32x^4+3875x^4y^4$\\
$(1,0,-1)$ & $x^4y^{-4}$\\
\hline
$(1,0,0)$ & $x^2y^{-2}+16x^3y^{-1}+136x^4$\\
\hline
$(2,0,0)$ & $x^4y^{-4}$\\
\hline
\caption{Contribution to the index from different GNO sectors in $U(3)_4$ theory.}
\end{longtable}

\begin{longtable}{|l|l|l|}
\hline
GNO charge & Top. charge & Index contribution\\
\hline
$0$ & 0 & $1+16xy+136x^2y^2+816x^3y^3-34x^4+x^4y^{-4}+3877x^4y^4$\\
\hline
 & 1 & $x^2y^{-2}-x^2y^{2}+16x^{3}y^{-1}-16x^{3}y^{3}+136x^4-136x^4y^4$\\
\hline 
& 2 & $-x^4+x^4y^{-4}$\\
\hline
$1$ & 0 & $x^4-x^4y^4$\\
 \hline
$-1$ & 0 & $x^4-x^4y^{4}$\\
 & 1 & $x^2y^2+16x^3y^3+136x^4y^4$\\
 & 2 & $x^4-x^4y^{4}+16x^3y-16x^3y^5$\\
 \hline
$-2$ & 2 & $x^4y^4$\\
\hline
\caption{Contribution to the index from different GNO sectors in $U(1)_4+M+v_{\pm}$ theory.}
\end{longtable}

\begin{longtable}{|l|l|}
\hline
GNO charge  & Index contribution\\
\hline
$0$ & $1+15xy-36x^2+105x^2y^2+21x^3y^{-1}-384x^3y+490x^3y^3$\\
\hline
  1 & $xy^{-3}$\\
\hline 
2 & $x^2y^{-6}$\\
\hline
3 & $x^3y^{-9}$\\
 \hline
\caption{Contribution to the index from different GNO sectors in $USp(2)_3$ theory.}
\end{longtable}

\begin{longtable}{|l|l|}
\hline
GNO charge & Index contribution\\
\hline
0 &  $1+xy^{-3}+15xy-xy^3+x^2y^{-6}-37x^2+120x^2y^2-15x^2y^4+x^3y^{-9}-x^3y^{-3}+$\\
& $36x^3y^{-1}-504x^3y+715x^3y^3-120x^3y^5$\\
\hline
 1 & $xy^3+x^2-15x^2y^2+15x^2y^4-x^2y^6+x^3y^{-3}-15x^3y^{-1}+120x^3y-226x^3y^3+$\\
& $135x^3y^5-15x^3y^7$\\
\hline 
2 & $x^2y^6+x^3y^3-15x^3y^5+15x^3y^7-x^3y^9$\\
\hline
3 & $x^3y^9$\\
 \hline
\caption{Contribution to the index from different GNO sectors in $USp(2)_3+15M+Y$ theory.}
\end{longtable}

\begin{longtable}{|l|l|}
\hline
GNO charge  & Index contribution\\
\hline
$(0,0)$ & $1+45xy-100x^2+1035x^2y^2+55x^3y^{-1}-4400x^3y+16005x^3y^3$\\
\hline
  $(1,0)$ & $xy^{-5}+45x^2y^{-4}-99x^3y^{-5}+825x^3y^{-3}$\\
\hline 
$(2,0)$ & $x^2y^{-10}+45x^3y^{-9}$\\
\hline
$(3,0)$ & $x^3y^{-10}$\\
 \hline
\caption{Contribution to the index from different GNO sectors in $USp(4)_5$ theory.}
\end{longtable}

\begin{longtable}{|l|l|}
\hline
GNO charge & Index contribution\\
\hline
$(0,0)$ &  $1xy^{-5}+45xy-xy^5-101x^2+x^2y^{-10}+45x^2y^{-4}+1035x^2y^2-45x^2y^6+$\\
& $x^3(y^{-10}+45y^{-9}-100y^{-5}+825y^{-3}+55y^{-1}-4445y+16215y^3+99y^5-1035y^7)$\\
\hline
 $(1,0)$ & $xy^5+x^2(1+45y^6-y^{10})+x^3(y^{-5}+45y-210y^3-100y^5+1035y^7-45y^{11})$\\
\hline 
$(2,0)$ & $x^2y^{10}+x^3(y^5+45y^{11}-y^{15})$\\
\hline
$(3,0)$ & $x^3y^{15}$\\
 \hline
\caption{Contribution to the index from different GNO sectors in $USp(4)_5+45M+Y$ theory.}
\end{longtable}

\section{Appendix B. Relations between generators of the chiral ring}

The fact that chiral operators $M_i^{\tilde\jmath}$ and $v_{\pm}$ are free generators of the chiral ring is obvious in the magnetic theory since there is no superpotential including both mesons and operators $v_{\pm}$.  In this appendix we prove this fact from the electric theory point of view not using duality.

If there is a relation between generators $M_i^{\tilde\jmath}$ and $v_{\pm}$ of the chiral ring, then there exist a monomial in these fields with zero topological, $U(1)_A$ charges and conformal dimension. This monomial has the expression $(v_-v_+)^nM^m=1$ where $n$ and $m$ are integral numbers, not necessarily positive. The condition of zero $U(1)$ charge is $-2N_fn+2m=0$. The equality to zero of the conformal dimension is equivalent to the condition of zero $R$-charge, which due to the condition on the $U(1)_A$-charge is just equality to zero of the UV $R$-charge $N_c=N_f+1$. This gives the conformal dimension of operators $v_{\pm}$: $\Delta(v_{\pm})=-\frac{N_f}{2}\Delta(M)$. Thus either the mesons or the minimal monopole operators $v_{\pm}$ have negative conformal dimension which violates unitarity. We conclude that there is no realtion between these operators.

\section{Appendix C. Consistency of the chiral ring}

 The purpose of this appendix is to show that for every bare monopole state in the deformed theory with GNO charges different from $(n,-m,0,...,0)$ there exists a possibility to become a part of a long supermultiplet and, correspondingly, become a $Q$-exact operator in the original theory. 

It was motivated in the main text that the only non-accidental BPS bare monopole states in the deformed theory are those with GNO charges $(n,-m,0,...,0)$ with nonnegative integral $n$ and $m$. A bare monopole with any other GNO charges must correspond to a $Q$-exact operator in the original theory. There are two scenarios how this can happen. The simplest one is that in the deformed theory for each bare monopole with GNO charges different from $(n,-m,0,...,0)$ there is a fermionic spinor state with quantum numbers appropriate for a $Q$-ascendant of the bare monopole. This cannot happen for bare monopoles $\ket{n,-m,0,...,0}$ because this is the lowest energy state in the sector with topological $U(1)_T$ charge $t=n-m$ and $U(1)_A$ charge $A=-N_f(|n|+|m|)$. In the second scenario there is not an appropriate fermionic superpartner for each bare monopole. But two conditions must be satisfied in order for the bare monopole to become $Q$-exact in the original theory. First, it cannot give a distinguished contribution to the index which unambiguously could be deciphered as that of a scalar BPS state. Second, there must be a mechanism explaining pairings of bare monopole states when the interactions are switched on. Below it is shown that the first scenario is not realized, but both conditions for the realization of the second scenario are met at least for several low values of $N_c$ and $N_f$.

 Consider some bare monopole state with GNO charges $(n_1,n_2,...,n_{N_c})$. The potential superpatner must be a state of the form $\Psi\ket{n,-m,0,...,0}$ where $\Psi$ is some monomial in the matter and gauge modes with spin one half. Moreover, $\Psi$ is the $SU(N_f)\times\widetilde{SU(N_f)}$ flavor singlet because all bare monopoles are flavor singlets. To get an idea how to build such a state in general, let us consider an example from the theory $U(3)_4$ in addition to those already discussed in the main text.

 Bare monopole $\ket{2,1,0}$. Its topological charge is $t=3$, $U(1)_A$-charge is $A=-12$ and the UV $R$-charge is $h=2$. 
The potential superpartner must be of the form $\Psi\ket{3,0,0}$, or $\Psi'\ket{4,-1,0}$ or $\Psi''\ket{5,-2,0}$, etc. The monomial $\Psi$ in the matter modes must be a singlet with respect to the flavor symmetry group $SU(4)\times\widetilde{SU(4)}$ because all bare monopoles are, so it is natural to look for an elementary monomial which is a singlet. These are $w\equiv\psi^iQ_i$ and $\tilde w\equiv\psi_{\tilde i}Q^{\tilde i}$. They have energy $3/2$ and spin $1/2$ as long as we take the lowest spin components of the matter scalars and the fermions. Moreover, these modes must not interact with nonzero magnetic charges. An obvious candidate for the fermionic state is $w_{+1/2}\ket{3,0,0}$ where the bare monopole $\ket{3,0,0}$ is chosen to have the same $U(1)_A$-charge and the UV $R$-charge as the state $\ket{1,1,1}$.

An important restriction on building a fermionic $Q$-ascendant of a bare monopole is that it must be a BPS state with $J_3=+1/2$. Indeed, the energy of this state is lower by one half, the $R$-charge is lower by one  and $J_3$ is higher by one-half. Thus $E-r-J_3=E_0-r_0=0$ where $E_0$ and $r_0$ are the energy and the $R$-charge of the bare monopole. As follows from the table with quantum numbers of the modes of all fields, all modes used to build a BPS spinor with $J_3=+1/2$ must be scalars with the exception only one which must have $J_3=j=1/2$. Moreover, $\Psi$ must be a flavor singlet. This means that we can use either $w_{+1/2}$ or $\tilde w_{+1/2}$ only once while the other modes must factorize into 'flavor baryons' and gauge-invariant scalar gluinos. Using this it is easy to show that many bare monopoles do not have appropriate fermionic states. Examples for the theory $U(3)_4$ are bare monopoles $\ket{2,2,-1}$ and $\ket{3,2,-2}$. 

The second scenario implies the two requirements whose satisfaction we show now.

\begin{itemize}

\item[{(a)}] No distinguished contribution to the index.

It was mentioned above that this requirement is not met for bare monopoles $\ket{n,-m,0,....,0}$ which sets them aside and guarantees their existence as BPS scalars in the original theory (put on $S^2\times{\mathbb R}$). 

\end{itemize}

For all other monopoles their contribution to the index, in principle, can be canceled by certain fermionic modes. For a bare monopole with topological $U(1)_T$ charge $t$ and $U(1)_A$ charge $A$ one available fermionic state is $w_{N/2}\ket{n,-m,0,...,0}$ where $n=\frac{t-A/N_f}{2}$, $m=-\frac{t+A/N_f}{2}$ and the mode $w_{N/2}$ is $\psi^iQ_i(s)$ where the $Q$-mode has spin $s$ determined from the requirement that the energy difference between the original bare monopole and $\ket{n,-m,0,...,0}$ is equal $2s+2$. Other fermionic states are obtained from different bare monopoles containing zero GNO charges.  We have been unable to show that for each bare monopole not of the type $\ket{n,-m,0,...,0}$ the contribution to the index is canceled by a fermionic state in general. With the increase in $N_c$ the numbers of 'unwanted' bare monopoles grow, but the number of compensating fermions grows as well, so it is not implausible that all contributions can be cancelled. We verified this for a number of low-energy monopoles for several low values of $N_f$ and $N_c$. 

One should note that all these modes $w$ and $\tilde w$ have even contributions to the value of $E+j_3$. So, for them to be useful, the energy difference between the bare monopoles must be even. This is always the case because, having equal $U(1)_A$ charges, their energy difference is determined by the difference in contributions coming from the vector multiplet $\delta E=\sum_{i<i}(|n_i-n_j|-|m_i-m_j|)$ which is always even for $\sum_in_i=\sum_im_i=t$. Moreover, for a given values of $U(1)_A$- and topological $U(1)_T$-charges the bare monopole $\ket{n,-m,0,...,0}$ has the lowest energy, which makes such state distinct.  Both statements are easy to prove by going from the initial bare monopole to the $\ket{n,-m,0,...,0}$ using a number of steps at each of which one of the GNO charges $n_i$ is increased by one while another $n_j$ is decreased by one without changing the $U(1)_A$. There is a sequence of such steps when the value of the expression $\sum_{i<j}|n_i-n_j|$ is increased by two at each step until the GNO charge $(n,-m,0,...,0)$ is reached.\footnote{The energy of a bare monopole $\ket{n_1,n_2,...,n_{N_c}}$ is given by the expression $E=-\sum_{i<j}|n_i-n_j|+N_f(1-\Delta)\sum_{i=1}^{N_c}|n_i|$. The $U(1)_A$ charge is $A=-N_f\sum_{i=1}^{N_c}|n_i|$.}

\begin{itemize}

\item[{(b)}] The pairing.

\end{itemize}

 The second condition necessary for $Q$-exactness of a scalar monopole operator is existence of a long multiplet near some value $t_0$ of the deformation parameter $t$ whose energy changes along the deformation so that at the point $t_0$ it breaks into short multiplets providing the bare BPS monopole with a $Q$-superpartner. There can be such multiplets in principle. An example of this is a long multiplet whose lower component is a spinor with energy satisfying the unitary inequality $E_s>r_s+j_s+1=r+3/2$. On the first level there is a scalar with energy $E_b>r_b+1$ and a vector. On the second level there is a spinor. At some point $t_0$ along the defrmation it may happen that $E_s=r_s+3/2$. In this situation the scalar from the first level with energy $E_b=r_b+1$ and spinor from the second level with energy $E=r+1/2$ become part of a separate short multiplet. This short multiplet has a zero-norm scalar state on the second level with energy $E=r$. Thus, the initial BPS scalar can take this place as the parameter of the deformation is varied further.   

The same mechanism can also govern the fate of the monopole operators defined in the asymptotically free UV theory along the RG flow. First, the free UV theory is put on $S^2\times {\mathbb R}$. Perturbation by the relevant operator of the theory on ${\mathbb R}^3$ that switches on the gauge interaction corresponds to turning on a time-dependent perturbation in the radially quantized picture in the far past. The nonunitary evolution leads to the radially quantized IR fixed point of the theory on ${\mathbb R}^3$ in the far future. Although this perturbation breaks time-translation invariance, the supersymmetry is preserved and states on the sphere $S^2$ are combined into supermultiplets. Initially, in the far past, the monopole operators live in short BPS multiplet, but when the interaction is switched on they can pair up with appropraite fermions into long multiplets. This can explain why most of the monopole operators may be absent in the chiral ring of the IR superconformal fixed point. Checking that the pairings actually occur is out of reach, but these pairings are possible in principle. The analysis of potential superpartners performed above for the deformed theory did not depend on any assumptions about values of anomalous dimensions. Thus, it is applicable to the case of canonical dimensions of all fields, and because the two analysises are identical, the picture is consistent.

\end{document}